\documentclass[fleqn,usenatbib,useAMS]{mnras}
\usepackage{graphicx}	
\usepackage{amsmath}	
\usepackage{amssymb}	
\usepackage{multicol}        
\usepackage{bm}		
\usepackage{pdflscape}	

\usepackage{mathrsfs}

\title[Magnetic White dwarfs : GR \& EoS]{Mass-Radius Relation of Strongly Magnetized White Dwarfs: Dependence on Field Geometry, GR effects and Electrostatic Corrections to the EOS}

\author[P. Bera and D. Bhattacharya]{Prasanta Bera\thanks{E-mail:pbera@iucaa.in}, Dipankar Bhattacharya\thanks{E-mail:dipankar@iucaa.in}\\
Inter University Centre for Astronomy and Astrophysics, Post Bag 4, Pune 411007, India.}

\begin{document}

\pagerange{\pageref{firstpage}--\pageref{lastpage}} \pubyear{0000}

\maketitle

\label{firstpage}

\begin{abstract}

Recent literature has seen an ongoing discussion on the limiting mass of strongly magnetized white dwarfs, since such objects may prove to be a source of over-luminous type-Ia supernovae. In an earlier paper, we have presented the mass-radius relation of white dwarfs with a strong poloidal magnetic field in Newtonian gravity. The inclusion of effects such as general relativistic gravity and many-body corrections to the equation of state can alter the mass-radius relation and the maximum mass. In this work we  estimate the extent to which these effects may modify the earlier results. We find that the general relativistic effects tend to reduce the maximum mass by about 2\% and many-body corrections by another additional $\sim$2\%, for an assumed carbon composition. We also explore field geometries that are purely toroidal or a mixture of poloidal and toroidal and find that the limiting mass of such equilibrium configurations can be substantially higher than in the case of a purely poloidal field.

\end{abstract}


\section{Introduction}

The inclusion of a strong internal magnetic field can substantially affect the structure and mass-radius relation of white dwarfs and raise their limiting mass. Such Super-Chandrasekhar mass white dwarfs could be possible progenitors of over-luminous type-Ia supernovae \citep{howel06,hicken_07,yamanaka_09, Scalzo+2010, Tanaka+2010, Silverman+2011, Taubenberger+2011}. Several publications in the recent literature have addressed different aspects of this issue: for example, the effects of magnetic field on the equation of state \citep{das_m12, Paret+15, Zou+Meng15, Mukhopadhyay+2015}, Lorentz force and instability \citep{nitya14, Coelho+14}, possibility of electron capture \citep{chame13, chamel_14, Vishal+Mukhopadhyay2014} and total lepton number violation \citep{belyaev15}. In an earlier work (hereafter P1, \cite{bera_b14}) we have presented the mass-radius relation of strongly magnetized white dwarfs with self consistent inclusion of Lorentz force. In this work, the gravitational field was treated in a 
Newtonian framework. Subsequently, \cite{das_m15} have pointed out that the inclusion of general relativistic (GR) effects may substantially reduce the upper mass limit. Here we revisit our earlier work with GR treatment of gravity and find that the reduction in the upper mass limit is at most $\sim$ 2.2\%.

Another important factor that affects the mass-radius relation of the white dwarfs is the equation of state (EoS) of the constituent material. In P1 we have used a pure Fermi degenerate electron EoS to describe the pressure density relation. However, as shown by \cite{salpeter61}, many body effects can have a significant effect on the EoS. In this work, we also explore the effects of many body corrections to the EoS on the mass-radius relation of magnetic white dwarfs. We find that for a white dwarf composed of carbon, this leads to a further reduction in the mass limit by about 2\%.

While the above discussion pertains to a purely poloidal configuration of the magnetic field, in reality the field configuration can be more complex. We have therefore also explored configurations with purely toroidal as well as mixed toroidal and poloidal fields. Our results in these cases show that the mass limit of equilibrium configurations could be substantially higher than that for pure poloidal configuration. While such a trend has been reported also by \cite{das_m15}, masses obtained by us lie well above the limits quoted by them.

In our work, we impose a constraint that the maximum of the mass density should occur at the center of the star. Recent work by \cite{Franzon+Schramm2015}, relaxing this assumption, finds that even higher mass equilibria may be obtained in a doughnut-shaped density distribution within the stellar interior.

In general, a white dwarf may possess fast rotation and this may affect its structure. However, we do not include the effects of rotation in the present work. The results presented here pertain to non-rotating white dwarfs with axisymmetric magnetic field distribution.

    The paper is structured as follows. In Section \ref{sol_method} we present in brief the method of solving the stellar structure equations and describe the equation of state (EoS) used by us. The results for different field configurations are presented in Section \ref{results}. Our conclusions are summarized in Section \ref{conclusions}.
 
\section {Equilibrium configurations with axisymmetric magnetic field} \label{sol_method}

To solve for the equilibrium configuration we use the spherical polar coordinates ($r$, $\theta$, $\phi$) with origin at the stellar center and the pole located along the axis of symmetry. We assume that the material has infinite conductivity, satisfying the ideal MHD condition with no electric field. The general approach we have followed to solve the stellar structure is modeled on the so-called ``Self Consistent Field method'' for both Newtonian and GR cases. These methods are briefly described in the subsections below. Further details on these methods may be found in \cite{hachi86, tomim05, lande09, bera_b14, bonazzola93, bocquet95, pili14}. To obtain the solution in Newtonian gravity, we use our earlier code \citep{bera_b14}, whereas we use publicly available codes $XNS$ \citep{Bucciantini11, pili14} and $Lorene$ \footnote{http://www.lorene.obspm.fr/} to obtain GR configurations. The XNS code finds self-consistently the axisymmetric equilibrium solution of the relativistic compact star with rotation and 
magnetic field assuming approximate gravity with a conformally flat 
condition (CFC) in GR 3+1 formalism. This code, capable of handling polytropic EoS, has been used to generate equilibrium structure of magnetized neutron stars with different field geometry \citep{pili14, pili15, bucciantini15}. It has also been used to study white dwarf configurations by approximating the Fermi degenerate EoS as a polytrope  \citep{das_m15, Subramanian+Mukhopadhyay2015}. We have modified the XNS code to use the full form of electron degenerate equation of state, thus avoiding the polytropic approximation. To obtain magnetized white dwarf configurations in full general relativistic gravity we use the {\em magstar} module of the spectral code {\em Lorene} which iteratively solves for the axisymmetric stellar configuration in 3+1 formalism. The output of {\em magstar} have been widely used to study stellar configurations and as input to the study of the evolution of poloidal fields \citep{liebling+10, gabler+12, gabler+13, haskell+14}. The advantage of using {\em magstar} is that it uses GRV2 
and GRV3 to test the accuracy of 
the result. GRV2 and GRV3 are similar to the virial condition in general relativity in 2d and 3d respectively \citep{bonazzola94, gourgoulhon94}.

\subsection{Newtonian Calculation}

In the presence of a magnetic field, there will be effects of Lorentz force on the stellar structure. In the Newtonian limit the stellar structure equations are
\begin{align}\label{hydro_eui}
\frac{1}{\rho}\mathbf{\nabla} P &=-\mathbf{\nabla} \Phi_g+\frac{1}{\rho}\left( \mathbf{j}\boldsymbol\times\mathbf{B}\right) \\
\mathbf{\nabla}^2\Phi_g &= 4\pi G \rho
\end{align}
\\where $P$, $\rho$, $\Phi_g$, $\mathbf{j}$ and $\mathbf{B}$ are pressure, mass density, gravitational potential, current density and  magnetic field respectively. In the ideal MHD condition, the magnetic field satisfies the following Maxwell's equations.
\begin{align}
 \nabla \cdot \mathbf{B} &= 0,  \\
 \nabla \times \mathbf{B} &= \mu_0 \mathbf{j} .
 \label{maxwell}
\end{align}
where $\mu_0$ is the free space permeability.

In axisymmetry, the set of equations ($\ref{hydro_eui}-\ref{maxwell}$) can be transformed, for Fermi degenerate EoS, to the following integral form using the Hachisu Self-Consistent-Field technique \citep{hachi86, tomim05, lande09}.

\begin{align}\label{intro_c}
 \dfrac{1}{\mu_{\rm e} m_{\rm B}} E_{\rm F} +\Phi_{\rm g} &=\mathcal M+C,
\end{align}
where $\mu_{\rm e}$ is the mean molecular weight per electron, $m_{\rm B}$ is the baryon mass, $E_{\rm F}$ is the Fermi energy, $\mathcal M$ is the function which expresses the magnetic influence on the structure and $C$ is a constant of integration. The functional form of $\mathcal M$ can be adjusted to give poloidal, toroidal or mixed field configurations. Equation ($\ref{intro_c}$) is used repeatedly at the grid points, along with proper boundary conditions, to converge on the stellar configuration iteratively.

For the poloidal and the mixed field geometry in axisymmetry, $\mathcal M$ is a function of the flux-function $u=r\sin\theta A_\phi$ where $A_\phi$ is the $\phi$ component of the magnetic vector potential. For simplicity, we assign to $\mathcal M$ a functional form proportional to $u$.
\begin{align}
 \mathcal M &= \rm{const.}\times \it{u},\\
 \mathbf B &=\frac{1}{r\sin\theta}\left( \boldsymbol\nabla u\boldsymbol\times \boldsymbol{\hat\phi}\right)+B_{\phi}\boldsymbol{\hat\phi}. \label{pol_B}
\end{align}
The $\phi$-component of the field $B_\phi$ is related to an arbitrary function of $u$: $B_\phi=f(u)/r\sin\theta$. Whereas for pure toroidal field $\mathcal M$ and $B_\phi$ are functions of $\rho r^2\sin^2\theta$, and these may be written as
\begin{align}
 \mathcal M &= -\frac{mK_m^2}{2m-1}(\rho r^2\sin^2\theta)^{2m-1}, \\
 B_\phi &= K_m\frac{(\rho r^2\sin^2\theta)^m}{r\sin\theta}, 	\label{tor_B}
\end{align}
here the index $m\geq 1$ and $K_m$ is a constant.

A stationary solution of the equilibrium structure is expected to satisfy the stellar virial condition, expressed as
\begin{equation}
 3\Pi+W+\mathscr{M}=0
\label{virial}
\end{equation}
\noindent
$$\text{here}\hspace{5mm}\Pi:\text{ contribution of internal energy}=\int PdV;$$ $$W:\text{ gravitational potential energy}=\frac{1}{2}\int\rho \Phi_{\rm g}dV;\hspace{2mm}$$\\and $$\mathscr{M}:\text{ magnetic energy}=\int\frac{B^2}{2\mu_0} dV; $$\\
where $V$ is the space volume.

The stellar configuration obtained from the self consistent field calculation is tested for the virial condition, the deviation being expressed in a non-dimensional form, 
\begin{equation}
\vert VC \vert=\frac{\vert 3\Pi+W+\mathscr{M}\vert}{\vert W\vert}.
\end{equation}
The solution is said to converge when the relative change of the configuration parameters fall below a pre-set limit. In all the configurations presented by us, the values of $\vert VC \vert$ are $\sim 10^{-5}$.
\subsection{General Relativistic Calculation}

To obtain the stellar configuration in the general relativistic case it is required to solve the Einstein equation
\begin{align}
R_{\mu\nu}-\frac{1}{2}g_{\mu\nu}R = \frac{8\pi G}{c^4} T_{\mu\nu}.
\end{align}
Here, the stress-energy tensor $T_{\mu\nu}$ contains matter and electromagnetic source terms
\begin{align}
T^{\mu \nu} = (e+P)u^\mu u^\nu + Pg^{\mu \nu} + \frac{1}{\mu_0}\left[F^{\mu\gamma}F^\nu_\gamma-\frac{1}{4}g^{\mu\nu}F_{\gamma\delta}F^{\gamma\delta}\right],
\end{align}
where $e$ is the total energy density, $\mathbf{u}$ is the fluid 4-velocity and $\mathbf{F}$ is the electromagnetic field tensor, which can be expressed using the potential 1-form $\mathbf{A}$ as:
\begin{align}
 F_{\mu\nu} = A_{\nu,\mu}-A_{\mu,\nu}.
\end{align}

$R_{\mu\nu}$ and $R$ are the Ricci tensor and Ricci scalar respectively. The metric tensor components $g_{\mu\nu}$ are used to express the line element as
\begin{align}
ds^2 = g_{\mu\nu}dx^{\mu}dx^{\nu}.
\end{align}
Following 3+1 formalism in axisymmetry the metric tensor can be expressed in spherical like coordinates $(t, r, \theta, \phi)$ as
\begin{align}
g_{\mu\nu}&dx^{\mu}dx^{\nu} = - N^2 dt^2 \notag\\
&+ A^4\left[B^2r^2\sin^2\theta(d\phi-N^\phi dt)^2 + \frac{1}{B^2}(dr^2+r^2d\theta^2)\right],
\label{Lorene_metric}
\end{align}
where $N (= e^\alpha)$, $N^\phi$, $A$ and $B$ are functions of ($r, \theta$) \citep{bocquet95}.

If the non-sphericity of the configuration is small, then the metric may be further assumed to be conformally flat in spatial dimension and the above expression approximated as \citep{pili14},
\begin{align}
g_{\mu\nu}dx^{\mu}dx^{\nu} &= - N^2 dt^2 \notag\\
&+ A^4\left[dr^2+r^2d\theta^2 + r^2\sin^2\theta(d\phi-N^\phi dt)^2\right].
\label{XNS_metric}
\end{align}

The projection of the energy-momentum conservation equation $T^{\mu \nu}_{~~;\mu}=0$ on time hyper-surfaces gives a 3-vector equation for the matter momentum density. The equation of stationary motion for perfect fluid with pure poloidal field can be expressed as:
\begin{align}
\frac{1}{e+P}\frac{\partial P}{\partial x^i} + \frac{\partial \alpha}{\partial x^i} = \frac{1}{e+P} j^\phi \frac{\partial A_\phi}{\partial x^i}
\label{GR_euler}
\end{align}
equation ($\ref{GR_euler}$) can be integrated to obtain an expression similar to that in equation ($\ref{intro_c}$),
\begin{align}
H + \alpha = \mathcal  M + \rm{const.}
\label{GR_C}
\end{align}
Here, $H$ is the heat function defined in terms of baryon number density $n$ as,
\begin{align}
H(n) = \int_0^n \frac{1}{e(n_1)+P(n_1)}\frac{dP}{dn}(n_1)dn_1.
\end{align}
where $e$ is the internal energy density.
Here too the functional form of $\mathcal  M$ decides the field configuration.

Equation (\ref{GR_C}) is used by the XNS and Lorene codes to find the stellar configuration. XNS uses the approximate line element form expressed in equation (\ref{XNS_metric}) whereas the full line element form of equation (\ref{Lorene_metric}) is used by the Lorene $magstar$ code. XNS has been designed to give solutions for poloidal, toroidal and mixed field conditions but $magstar$ handles only a pure poloidal field geometry. Pure toroidal configurations using the full metric of equation (\ref{Lorene_metric}) can be similarly obtained \citep{Kiuchi+Yoshida2008, Frieben+Rezzolla2012}, but we have not attempted this in the present work. The functional form of the magnetic term $\mathcal  M$ required to model a pure poloidal or a pure toroidal field is similar to that mentioned in the Newtonian case.

\subsection {Equation of State (EoS)}
White dwarfs are dense objects and their interior temperature is small compared to the Fermi temperature ($T_F$) of electrons. So the EoS is very similar to that of a zero temperature Fermi electron gas, of which the relation between the baryon density ($\rho$) and pressure ($P$) may be expressed as follows:
\begin{align}
n_{\rm e} &=  \frac{1}{3\pi^2\lambda_{\rm e}^3}x_{\rm F}^3\\
\rho &=\mu_{\rm e} m_{\rm B}n_{\rm e}	\\
e &= \frac{\pi m_e^4c^5}{h^3}\left[ x_{\rm F}(1+2x_{\rm F}^2)\sqrt{1+x_{\rm F}^2}-\sinh^{-1}x_{\rm F}\right] \\
P &=\frac{\pi m_e^4c^5}{3h^3}\left[ x_{\rm F}(2x_{\rm F}^2-3)\sqrt{1+x_{\rm F}^2}+3\sinh^{-1}x_{\rm F}\right].
\end{align}
Here, $n_{\rm e}$ is the electron number density, $\lambda_{\rm e}=\frac{\hbar}{m_{\rm e}c}$, $x_{\rm F}=\frac{p_{\rm F}}{m_{\rm e}c}$, $p_{\rm F}$ being the Fermi momentum, $e$ is the internal energy density, $h$ is the Plank constant, $\hbar$ = $h/2\pi$ and $c$ is the speed of light.

Corrections to the above equation of state may arise from a number of effects. Coulomb interaction in the ion lattice is one source of such corrections. \cite{salpeter61} considered corrections to the pressure and energy expressions from i) Coulomb energy due to ion lattice, ii) Thomas-Fermi effect due to non-uniform charge distribution, iii) spin-spin interaction and exchange energy between electrons and iv) correlation energy term. At very high densities the electron Fermi energy becomes high enough to enable neutronisation to occur via inverse beta decay. The 
Neutronisation Threshold for He and C are 20.596 and 13.37 MeV respectively \citep{shapi83}. Beyond this the configuration becomes unstable. \cite{hamada_sol61} have presented the mass radius relation of non-magnetic white dwarfs, including these corrections. We investigate in this work the effect of these same corrections on the mass-radius relation of strongly magnetized white dwarfs. The relativistic Feynman-Metropolis-Teller EoS \citep{rotondo+11} is an improvement over the Salpeter one, but yields corrections very similar to the Salpeter EoS. 

The strong magnetic field may itself modify the equation of state of white dwarf matter by trapping electrons in quantized Landau levels. The effects of this are most severe when electrons inhabit only the lowest Landau levels. However, the condition in a realistic stellar configuration always distributes electrons over several tens of levels, minimizing such impact \citep{bera_b14}. Strong internal magnetic fields may affect the many-body interactions and modify the Salpeter corrections, however, we have not included the effect of Landau quantization and magnetic modifications of the many-body interactions in the work presented here.

\section{Results}\label{results}
In the following sections, we present the results obtained using the methods outlined in the previous section. Initially different magnetic field configurations from the Newtonian calculation are constructed and then we compare pure poloidal and toroidal field configurations with their GR results. Next we present the effects of inclusion of the ion interactions in the EoS for poloidal field configurations, in general relativistic gravity. We use the baryon mass to represent the mass of the configuration and adopt a value of $1.9885\times10^{30}$ kg for the solar mass unit (M$_\odot$). In all GR configurations, explored by us, the gravitational mass differs from its baryon mass by less than $10^{-3}$. In representing the mass-radius relation of non-spherical configurations, we use either the polar or the equatorial radius, whichever is larger.

\subsection {Magnetic configurations for different field geometry : Newtonian calculation}

\subsubsection{Poloidal field}
\begin{figure*}
\centering
\includegraphics[width=0.95\textwidth]{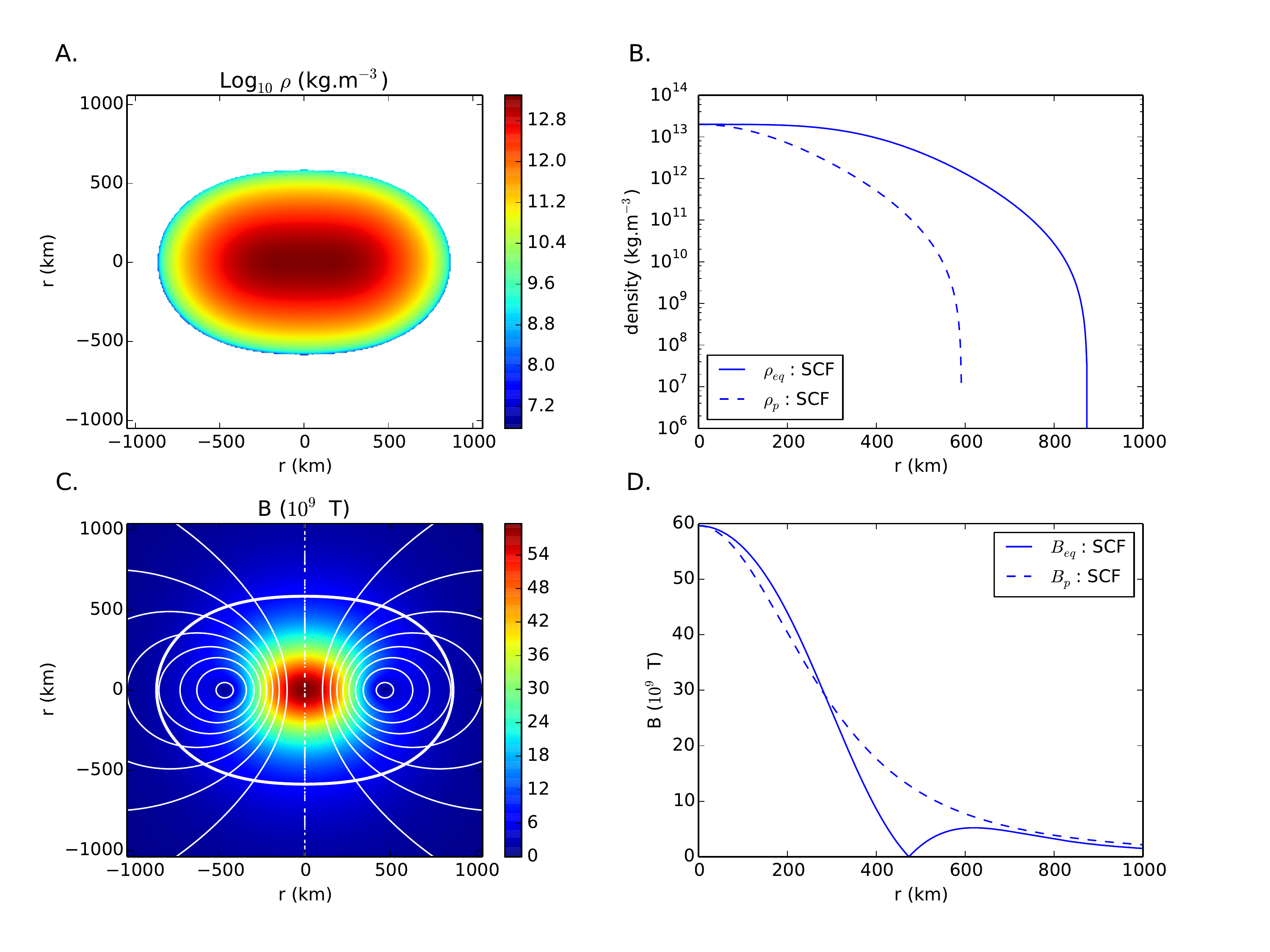}
\includegraphics[width=0.65\textwidth]{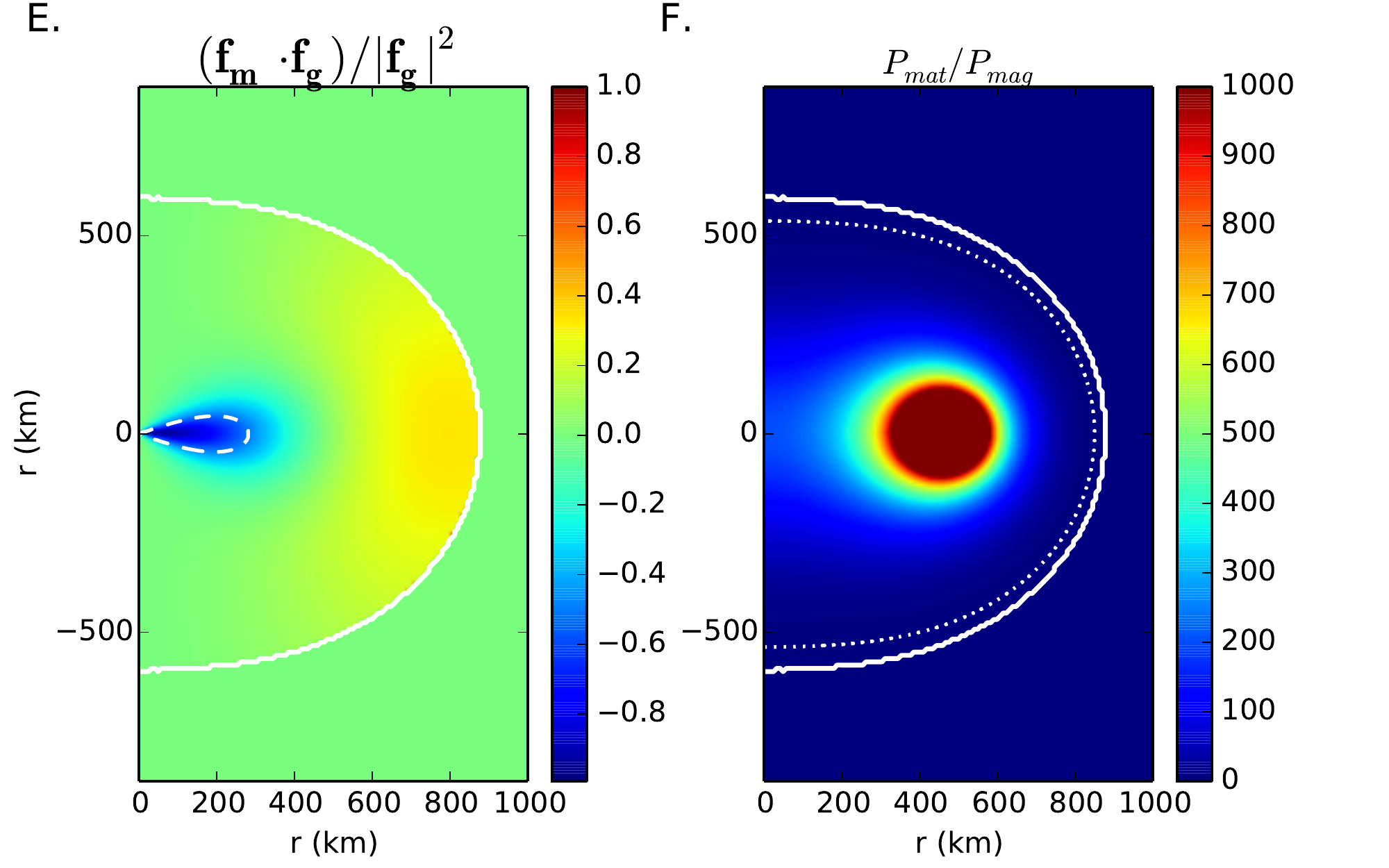}
\caption{The distribution of mass density and magnetic field strength inside a white dwarf configuration with a strong poloidal field from the Newtonian SCF calculation. The stellar outer boundary is indicated by a thick line. \textbf{A.} The density distribution shows an oblate structure, \textbf{B.} The radial profile of mass density along polar and equatorial directions, \textbf{C.} The 2-d poloidal field distribution, \textbf{D.} The radial distribution of magnetic field strength along polar and equatorial directions. \textbf{E.} The Lorentz force component along the gravitational force, the dashed white contour marking the region where the Lorentz force is 90\% in magnitude of the gravitational force, and opposite in direction. \textbf{F.} The distribution of plasma beta, the ratio of matter pressure to magnetic pressure in this configuration, the dotted contour indicating the value of 0.5. Plasma beta is seen to be high through most of the interior. }
\label{poloidal_den}
\end{figure*}

In our calculations, the poloidal field is specified through a current distribution which is purely toroidal. The functional form of the toroidal current distribution is expressed  as $J_\phi = r\sin\theta \rho \alpha_0$. The parameter $\alpha_0$ describes the strength of the current. An example of a resulting configuration is shown in Fig. \ref{poloidal_den}. This object has a central density $2\times10^{13}\rm{kg.m^{-3}}$ and a central field strength 59.6$\times10^9$ T. The Newtonian calculation gives a mass of 1.896 M$_\odot$ and a radius of 875.3 km for this object. A star with such a strong poloidal field has an oblate shape. The ratio of polar to equatorial radius in this case is about 0.68 and the ratio of magnetic to gravitational energy is about 13\%. Near the equatorial plane in deep interior of the star the Lorentz force rises to become almost equal and opposite to the gravitational force. However the plasma beta remains high throughout most of the stellar configuration.

\begin{table*}
 \begin{tabular}{c|c|c|c}
  \hline
    \bf        		&	\bf{SCF}		&	\bf{XNS}		&	\bf{Lorene}	\\
  \hline \hline
	\bf CONDITIONS	&				&			&\\
	
	grid 		&	512$\times$512		&	 \begin{tabular}[c]{@{}c@{}} NR = 500, NTH = 100\\($R_{max}=2500$)\end{tabular}	&	\begin{tabular}[c]{@{}c@{}} nr = 129 (inside star) nt = 65\\ nr = 65  (outside star) nt = 65\end{tabular}	\\
	
	$\rho_c$	&	2$\times10^{13}$ kg.m$^{-3}$&	2$\times10^{13}$ kg.m$^{-3}$&	2$\times10^{13}$ kg.m$^{-3}$	\\
	Field parameter	&	$B_{core}$ = 59.621 GT	&	$k_{pol}$ = 0.01164, $\xi$ = 0	&	CFA = 950 \\
	\hline
	
	\bf   RESULTS	&				&\\
	Mass (M$_\odot$)&	1.8965			&  	1.8557  	 	&  	1.8540		\\
	R$_{eq}$ (KM)	&	875.26			&  	860.16	  		&  	865.95		\\
	R$_p$/R$_{eq}$	&	0.6758			&	0.6738			&	0.6767		\\
	B$_{max}$ (GT)	&	59.621 			&	59.893			&	59.621		\\
	$|\mathcal{M}/W|$	&	0.1288		&	0.1305			& 	0.1276		\\
	Virial test	&	$\mid$VC$\mid$ = 7.84$\times10^{-6}$	&	-			&	\begin{tabular}[c]{@{}c@{}} $\mid$GRV2$\mid$ = 2.98$\times10^{-8}$\\ $\mid$GRV3$\mid$ = 1.56$\times10^{-8}$\end{tabular}	\\
  \hline
 \end{tabular}
\caption{Pure poloidal field configurations from Newtonian SCF calculation and from GR codes XNS and Lorene.}
\label{poloidal_XNS_Lorene_SCF}
\end{table*}

For a given central density and field geometry, a higher field strength leads to a larger mass of the equilibrium configuration. The mass-radius relation of strongly magnetized white dwarfs is thus dependent on the field parameters (as shown in Fig. 4. of P1 or Fig. $\ref{poloidal}$). Here we have used the central field strength or the ratio of magnetic field energy to gravitational energy to represent the field parameter. For a fixed central field strength, as the central density is reduced the magnetic energy becomes more dominant, causing the mass-radius relation to deviate strongly from that of non-magnetic configurations. We present here solutions in which the component of Lorentz force just opposite to gravity never exceeds in magnitude the gravitational force itself. Violation of this would move the density maximum away from the stellar centre, as shown by \cite{Franzon+Schramm2015}.
\subsubsection{Toroidal field}

\begin{figure*}
\centering
\includegraphics[width=0.95\textwidth]{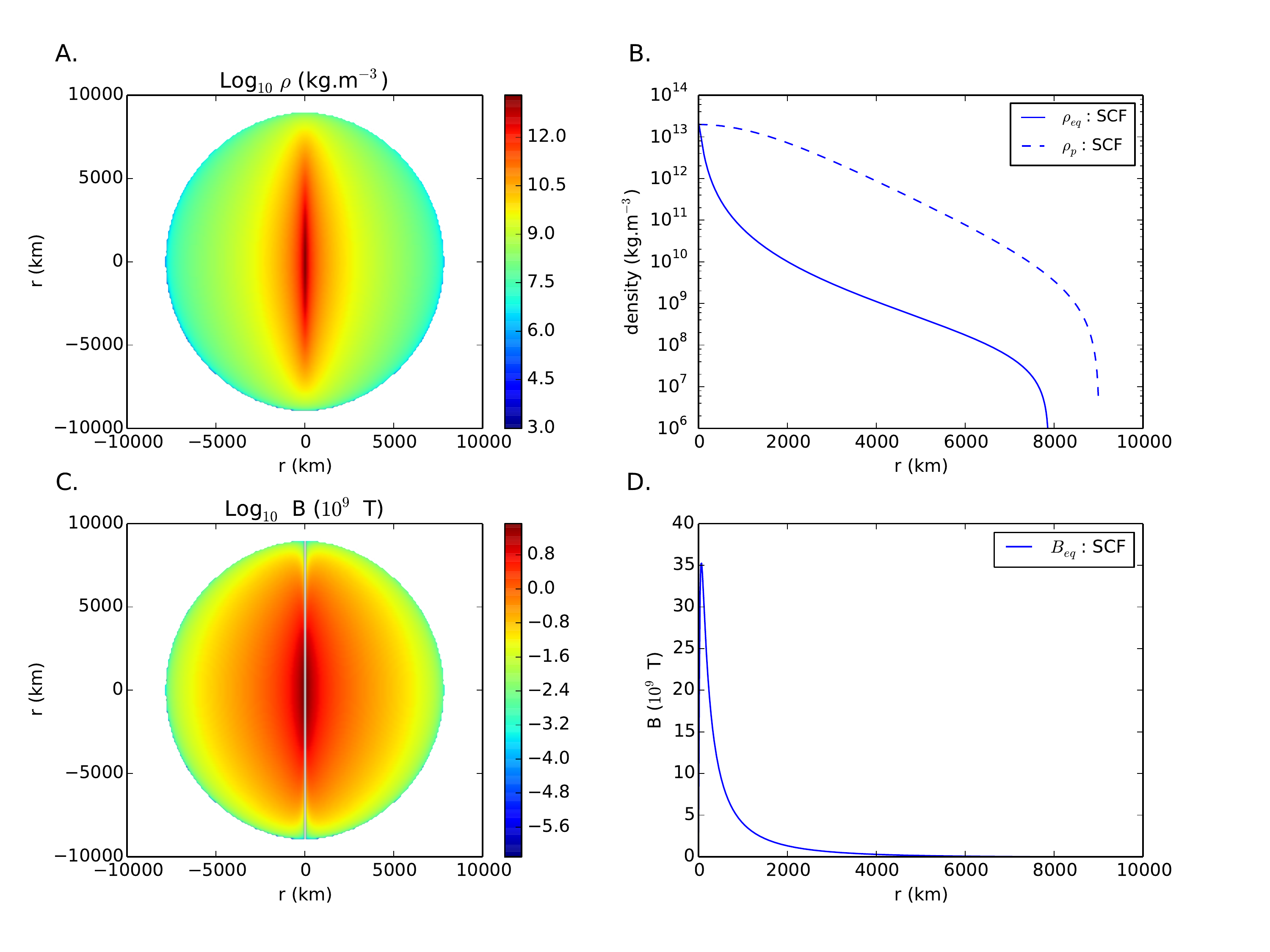}
\includegraphics[width=0.65\textwidth]{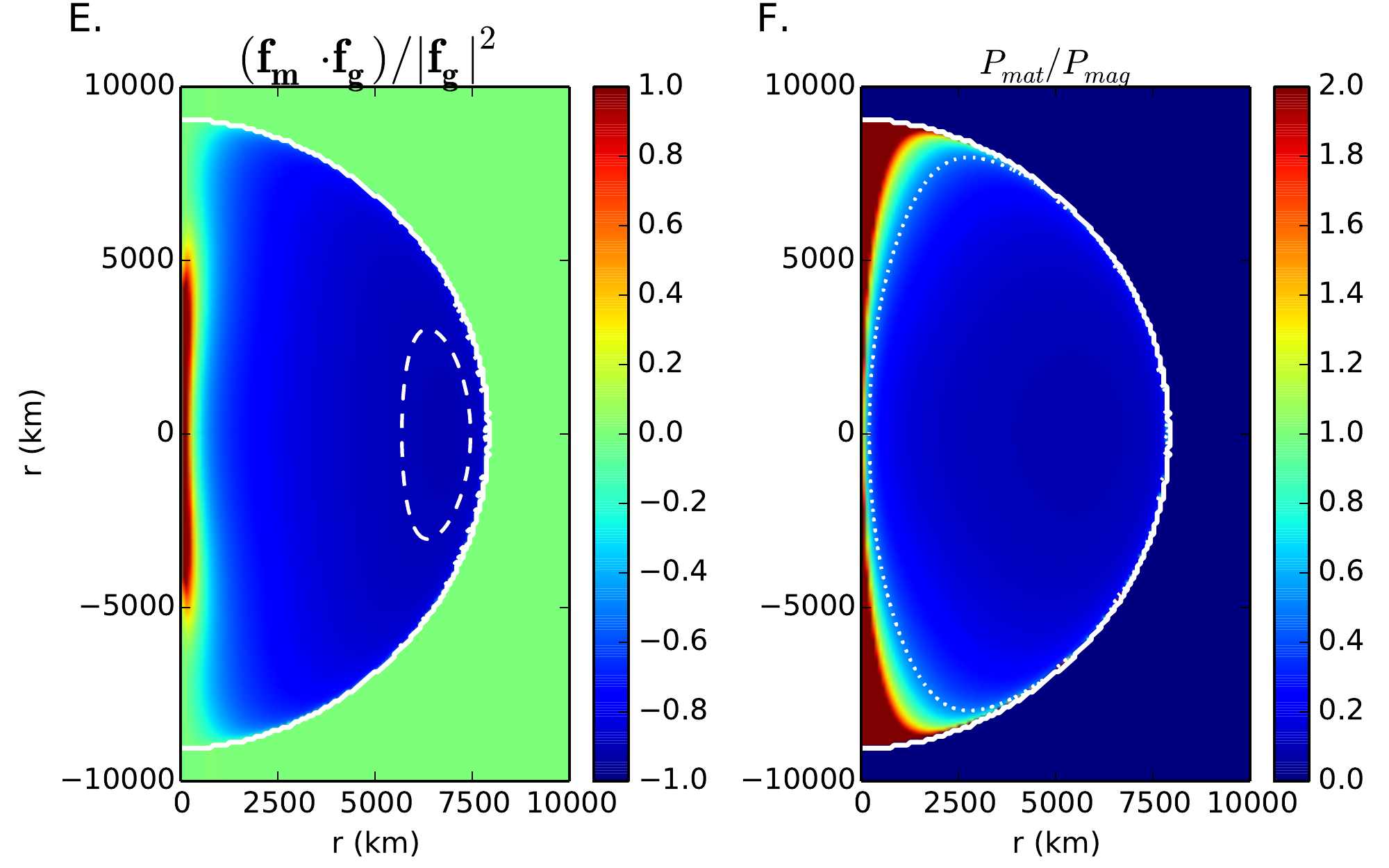}
\caption{Same as Fig. \ref{poloidal_den} but with pure toroidal field. \textbf{A.} The density distribution shows that the structure is prolate. \textbf{B.} The radial density profile along polar and equatorial directions. \textbf{C.} The 2-d toroidal field distribution. \textbf{D.} The equatorial field strength profile. \textbf{E.} The Lorentz force component along the gravitational force, the white dashed contour indicating the region where gravity is being opposed by a Lorentz force component that is 90\% as strong. \textbf{F.} Distribution of plasma beta inside the configuration, the dotted contour indicating a value of 0.5. The beta values are much lower in this configuration than in the poloidal case.}
\label{toroidal_2d}
\end{figure*}

A star with a pure toroidal field ($\mathbf{B} = B_\phi\hat\phi$) is prolate and the field is fully confined within the star with zero surface field strength \citep{fujisawa+eriguchi15}. We model such a magnetic field distribution by setting $m=1$ in equation ($\ref{tor_B}$), i.e. $B_\phi=K_m r \sin\theta\rho$, where $K_m$ is a constant. Pure toroidal field configurations have zero field at the center and along the axis of symmetry. Therefore in this case we use the total magnetic energy as the representative magnetic field parameter. A typical stellar configuration arising from such a field distribution is shown in Fig. $\ref{toroidal_2d}$. In this example the central density is 2$\times10^{13}$ kg.m$^{-3}$ and the ratio of magnetic energy to gravitational energy ($|\mathcal{M}/W|$) is about 39\%.
\begin{table*}
 \begin{tabular}{c|c|c}
  \hline
  \bf        		&	\bf{SCF}			&	\bf{XNS}	\\
  \hline \hline
	\bf CONDITIONS	&					&			\\
		grid	&	 1024$\times$1024		&	\begin{tabular}[c]{@{}c@{}}NR = 1000, NTH = 250\\ ($R_{max}=10000$)\end{tabular}\\
		$\rho_c$&	2$\times10^{13}$ kg.m$^{-3}$	&	2$\times10^{13}$ kg.m$^{-3}$	\\
	Field parameter	&	$|\mathcal{M}/W|$ = 0.3928, m = 1		&	$K_m$ = 7, m = 1\\
	\hline
	\bf   RESULTS	&\\
	Mass (M$_\odot$)&  	5.1611  	 		&  	4.9490	\\
	R$_{eq}$ (KM)	&	7854.6 		  		&	7774.6	\\
	R$_{p}$/R$_{eq}$&	1.1441				&	1.1386		\\
	$|\mathcal{M}/W|$	&	0.3928			&  	0.3928	\\
	virial condition&	$\mid$VC$\mid$ = $6.27\times10^{-5}$	&	-	\\
  \hline
 \end{tabular}
\caption{Pure toroidal field configurations from SCF Newtonian calculation and from XNS code.}\label{toroidal_XNS_SCF}
\end{table*}

For strong toroidal fields, highly massive stellar configurations are obtained as the Lorentz force is able to reduce the effective inward gravity for most of the interior (see Fig. $\ref{toroidal_2d}$ \textbf{F}). The value of plasma beta is less than unity almost everywhere except a small region near the axis. Such configurations are likely to be  strongly prone to interchange instabilities. 

\begin{figure}
\centering
\includegraphics[width=0.47\textwidth]{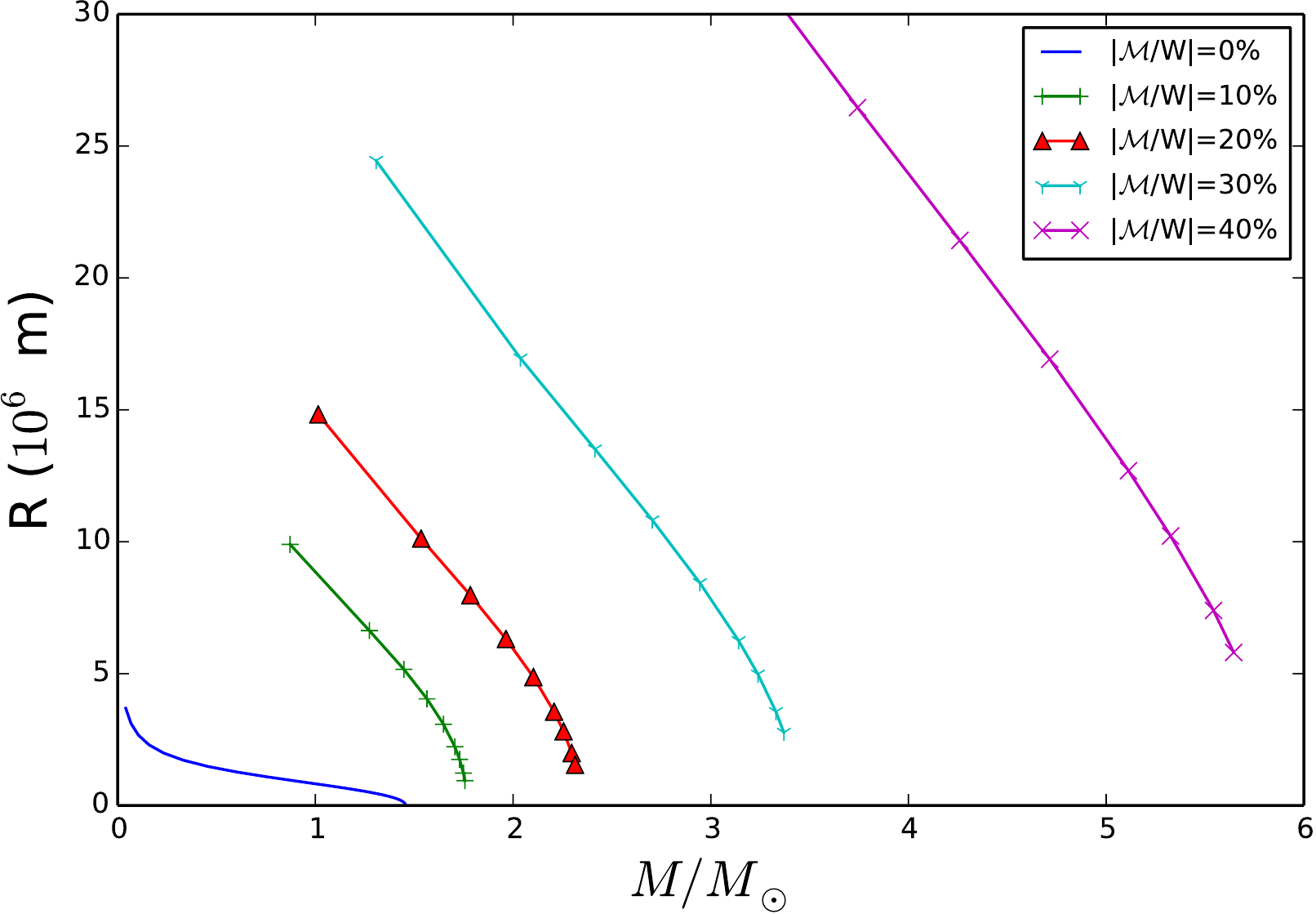}
\caption{Mass-radius relation of white dwarfs with strong toroidal magnetic field, as derived from Newtonian SCF calculations. Curves for different values of $|\mathcal{M}/W|$ are shown. Masses here can far exceed the traditional Chandrasekhar limit.}
\label{toroidal_M_R}
\end{figure}

Fig.~$\ref{toroidal_M_R}$ displays the mass radius relation for white dwarfs with pure toroidal field in the interior, as obtained from Newtonian SCF calculations.  Several curves for different values of  $|\mathcal{M}/W|$ are shown.  As the field energy increases, both the mass and the radius of the configuration increase in comparison to   non-magnetic configurations. As seen in the figure, for $|\mathcal{M}/W|=10\%$ the maximum mass is about 1.76 M$_\odot$, while for $|\mathcal{M}/W|=40\%$ the maximum mass reaches a value as high as 5.6 M$_\odot$.

\subsubsection{Mixed field}

\begin{figure*}
\centering
\includegraphics[width=0.95\textwidth]{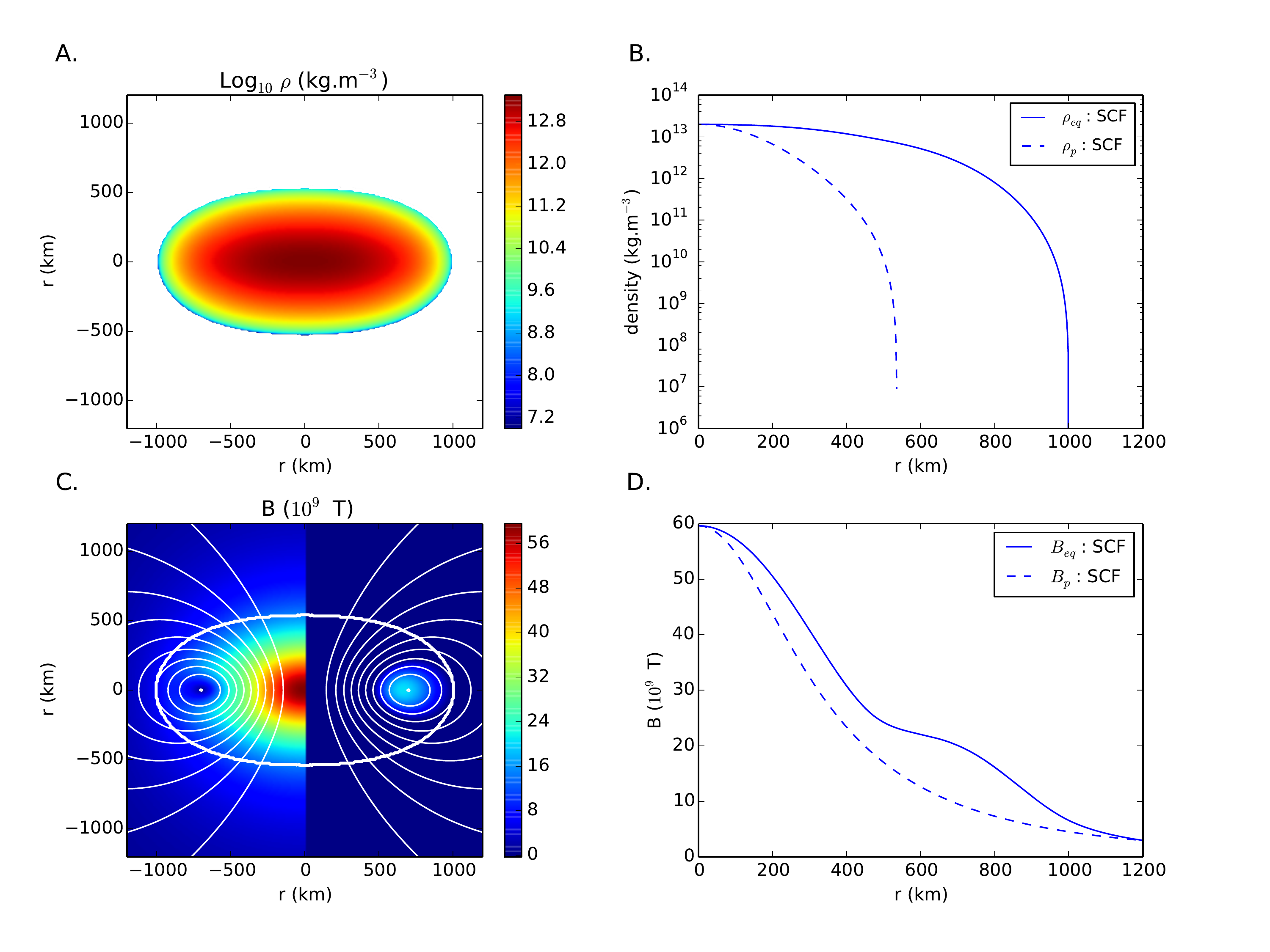}
\includegraphics[width=0.65\textwidth]{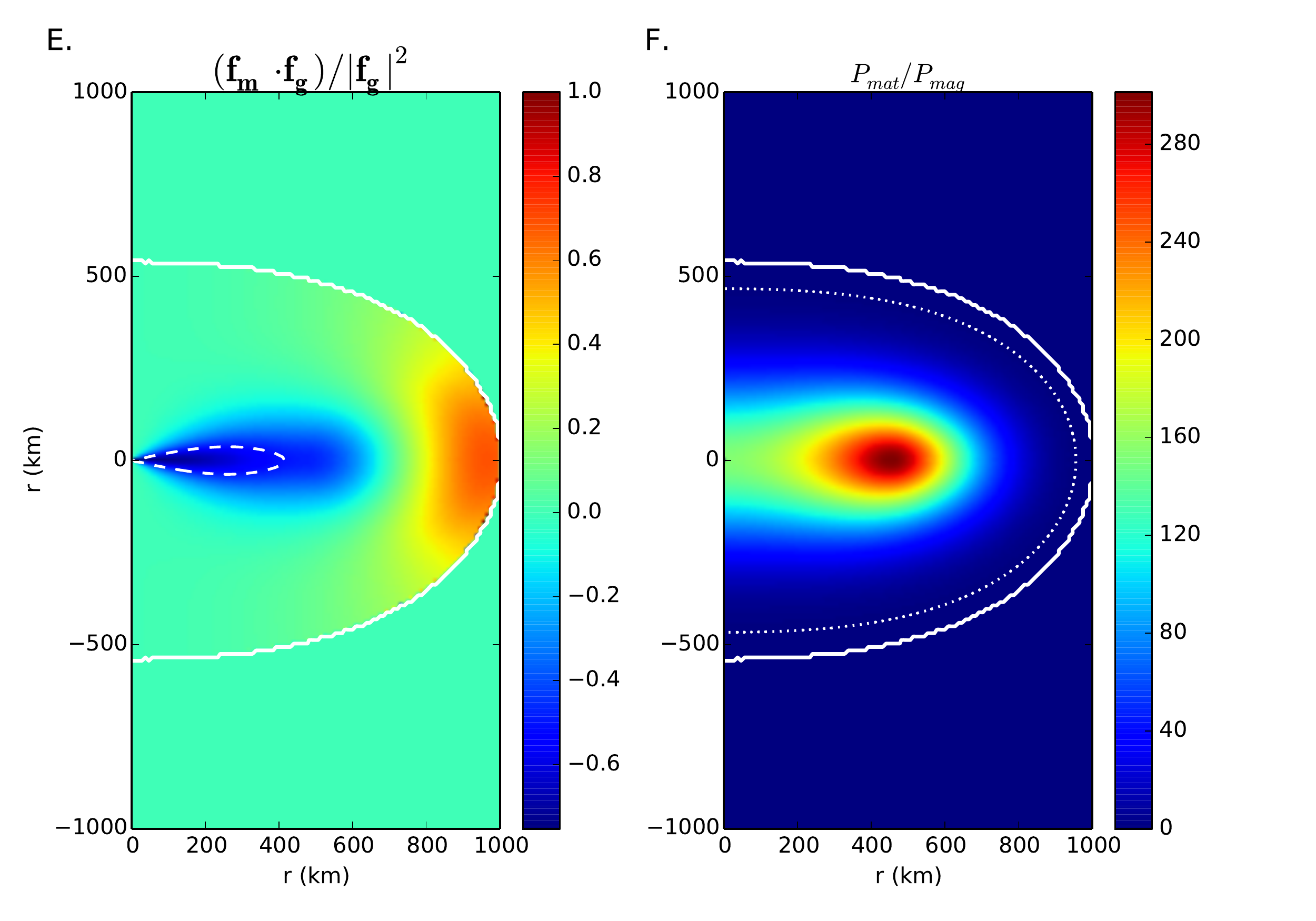}
\caption{Same as Figure~\ref{poloidal_den} but for a mixed poloidal and toroidal field, as in configuration conf.~II of Table \ref{mixed_SCF}. \textbf{A.} The density distribution shows a highly oblate structure. \textbf{B.} The radial distribution of mass density along polar and equatorial directions. \textbf{C.} The field distribution in the interior: the poloidal and toroidal components are plotted separately in the left and the right halves respectively. The white contours represent the field lines. \textbf{D.} The polar and equatorial distribution of the field strength. \textbf{E.} The Lorentz force opposing gravity rises to a maximum of about 75\% in this case, the white dashed contour indicating 50\%.  \textbf{F.} Plasma beta value is very high over most of the star. The dotted contour indicating plasma beta value of 0.5 is located close to the stellar surface.}
\label{mixed_2d}
\end{figure*}

In order to investigate structures with a mixture of poloidal and toroidal fields, we use a functional form $f(u)=\frac{\kappa_0}{k+1}(u-u_{max})^{k+1}$, in equation (\ref{pol_B}). We set $k=0$ and $\mathcal M = \alpha_0 u$ and carry out a Newtonian SCF calculation. Comparing with a pure poloidal field structure of the same central density and central field strength, we find that the introduction of a toroidal field increases both the total mass and the oblateness of the configuration. As listed in Table \ref{mixed_SCF}, the mass increases from 1.896 M$_\odot$ to 2.11 M$_\odot$ and then to 2.55 M$_\odot$ as the fraction of the field energy in the toroidal component is increased from zero to 4.7\% and to 7\%. The corresponding oblateness also increases, with R$_p$/R$_{eq}$ going from 0.68 to 0.62 and 0.53 respectively. The field distribution shows that the toroidal field is confined near the stellar surface with its strength reaching a maximum where the poloidal field attains zero magnitude (Fig. \ref{mixed_2d}). 
The values of plasma beta remain high in the interior parts, indicating a relatively stable structure.

\begin{table*}
 \begin{tabular}{|c|c|c|c|}
  \hline
    \bf        	Code : SCF	&	\bf{Mixed field conf. I}	&	\bf{Mixed field conf. II}\\
  \hline \hline
  	\bf CONDITIONS	&			&\\
	$\rho_c$	&	2$\times10^{13}$ kg.m$^{-3}$&	2$\times10^{13}$ kg.m$^{-3}$\\
	Field parameter	&	B$_{core}$ = 59.621, $\hat{\kappa_0}$ = 5 	&	B$_{core}$ = 59.621 GT, $\hat{\kappa_0}$ = 8		\\
	\hline
	\bf   RESULTS	&				&			\\
	Mass (M$_\odot$)&  	2.1128		&	2.5552	\\
	R$_{eq}$ (KM)	&  	913.32		&	1000.7	\\
	R$_p$/R$_{eq}$	&	0.6211		&	0.5352	\\
	B$_{max}$ (GT)	&	59.621		&	59.621	\\
	$|\mathcal{M} /W|$	&	0.1797	&	0.2612	\\
	$|\mathcal{M}_{tor} /\mathcal{M}|$	&	4.6753$\times10^{-2}$	&	7.0047$\times10^{-2}$	\\
	Virial test		&	$\mid$VC$\mid$ = 2.29$\times10^{-5}$	&	$\mid$VC$\mid$ = 5.39$\times10^{-5}$	\\
  \hline
 \end{tabular}
\caption{Mixed field configurations from Newtonian SCF calculation with different toroidal field parameter $\hat{\kappa_0}$ for a fixed central mass density and central field strength. Here $\hat{\kappa_0}$ is a non-dimensional constant that appears in the function $f(u)$.}
\label{mixed_SCF}
\end{table*}

\subsection{GR effects on white dwarf structure}

In order to quantify the effect of General Relativity on the mass and radius of a magnetic white dwarf, we intend 
to compare the configurations obtained using Newtonian and GR methods for the same central density and magnetic field strength. However this is not straightforward  because the field strength cannot be directly specified in XNS and Lorene codes as input, instead other custom parameters are used. The field distribution obtained can only be quantified in the output solution. Therefore, an iterative adjustment of input parameters to these codes is required to obtain the desired interior field strength. We compare here solutions of these three codes when their field strengths differ by less than 1\%, for the same value of central density.

\subsubsection {Non-magnetic configuration: comparison of XNS, Lorene and TOV results}

\begin{table*}
 \begin{tabular}{c|c|c|c|c|c}
  \hline
  \bf        &  TOV  &  XNS &  relative difference (XNS \& TOV)  &  Lorene  & relative difference (Lorene \& TOV)	\\
  \hline
  \bf  Mass  &  1.4158 M$_\odot$  &  1.4155 M$_\odot$	&  0.02\%  &  1.4167 M$_\odot$ & 0.06\% \\
  \bf  Radius&  1075 km  &	1067 km &  0.74\% &  1072 km & 0.3\%	\\
  \hline
 \end{tabular}
\caption{Non-magnetic configurations with Fermi degenerate EoS. Here central density = 2$\times10^{13}$ kg.m$^{-3}$, XNS grid : NR = 500, NTH = 100, Lorene grid : nr = 129 nt = 65 (inside star) \& nr = 65 nt = 65 (outside star).}\label{TOV_XNS}
\end{table*}

We have modified the GR codes XNS and Lorene to use the full form of the Fermi degenerate EoS for the white dwarf structure. To validate the modifications of these codes, we first compare the results for non-magnetic spherically symmetric configurations with those obtained from the Tolman-Oppenheimer-Volkoff (TOV) equation \citep{tolman39,oppenheimer+volkoff39}.

\begin{figure}
\centering
\includegraphics[width=0.47\textwidth]{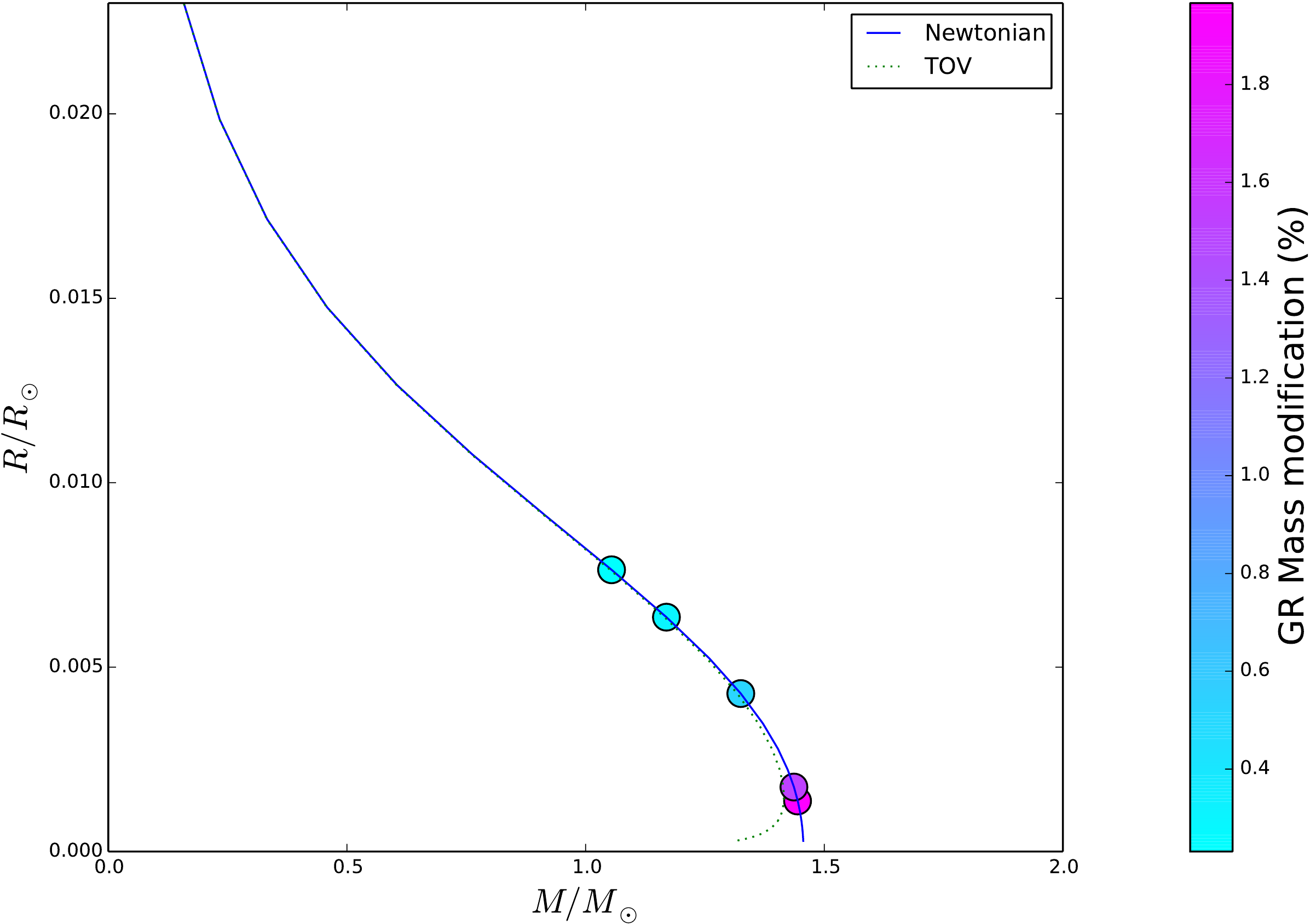}
\caption{Comparison of Newtonian and GR estimates of masses and radii of non-magnetic White Dwarfs.  The maximum mass values are 1.45 M$_\odot$ and 1.41 M$_\odot$ in Newtonian and GR calculations respectively. The circles represent the percentage of mass modification upon inclusion of GR effects.}
\label{TOV_MR}
\end{figure}

The mass-radius relation of white dwarfs resulting from the TOV equation is compared with that from Newtonian calculations in Fig. $\ref{TOV_MR}$. For low mass and larger radius the GR mass-radius curve closely follows the Newtonian results, but for very compact configurations the curve turns around to smaller masses, indicating gravitational instability. To quantify the effect of GR, we compare the masses derived for the same central density in Newtonian and GR calculations. The relative difference between these values show that the change in mass due to GR effects is less than 2\% for the stable configurations (see Fig. $\ref{TOV_MR}$).

\begin{figure}
\centering
\includegraphics[width=0.47\textwidth]{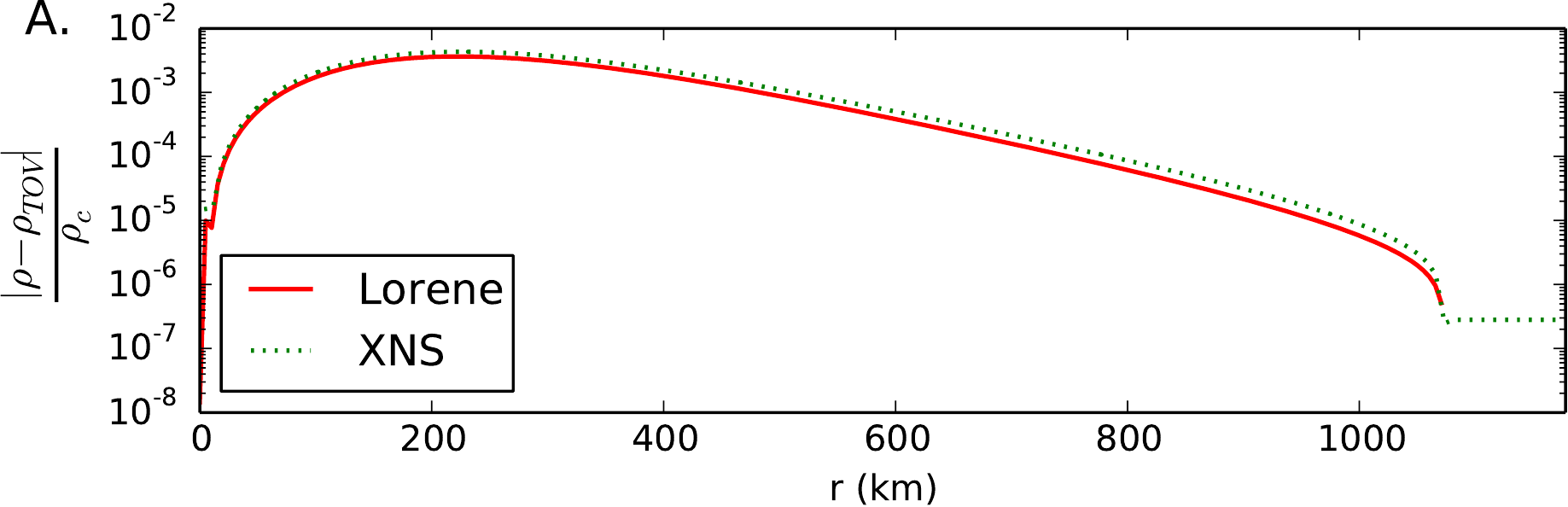}
\includegraphics[width=0.47\textwidth]{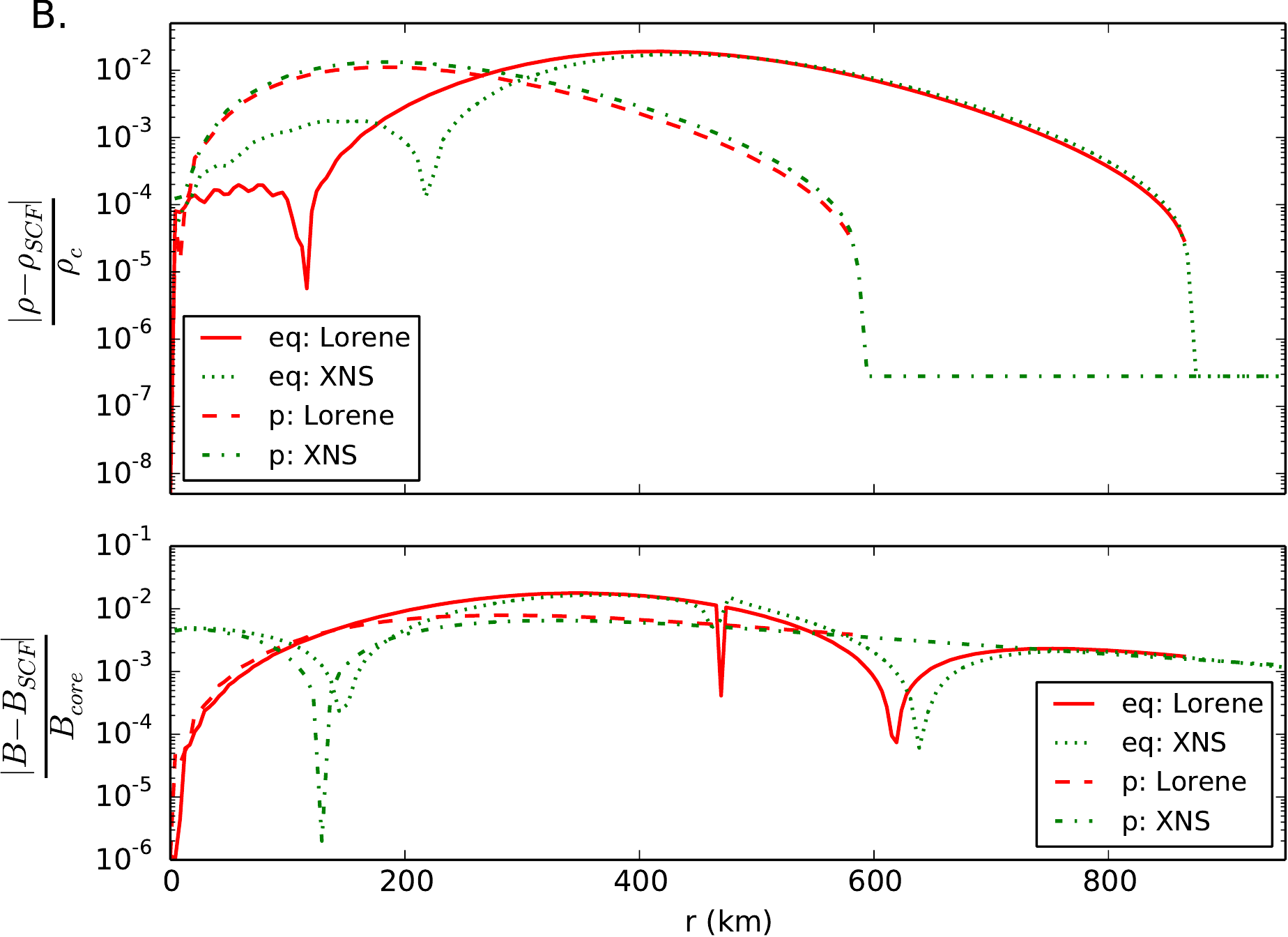}
\includegraphics[width=0.47\textwidth]{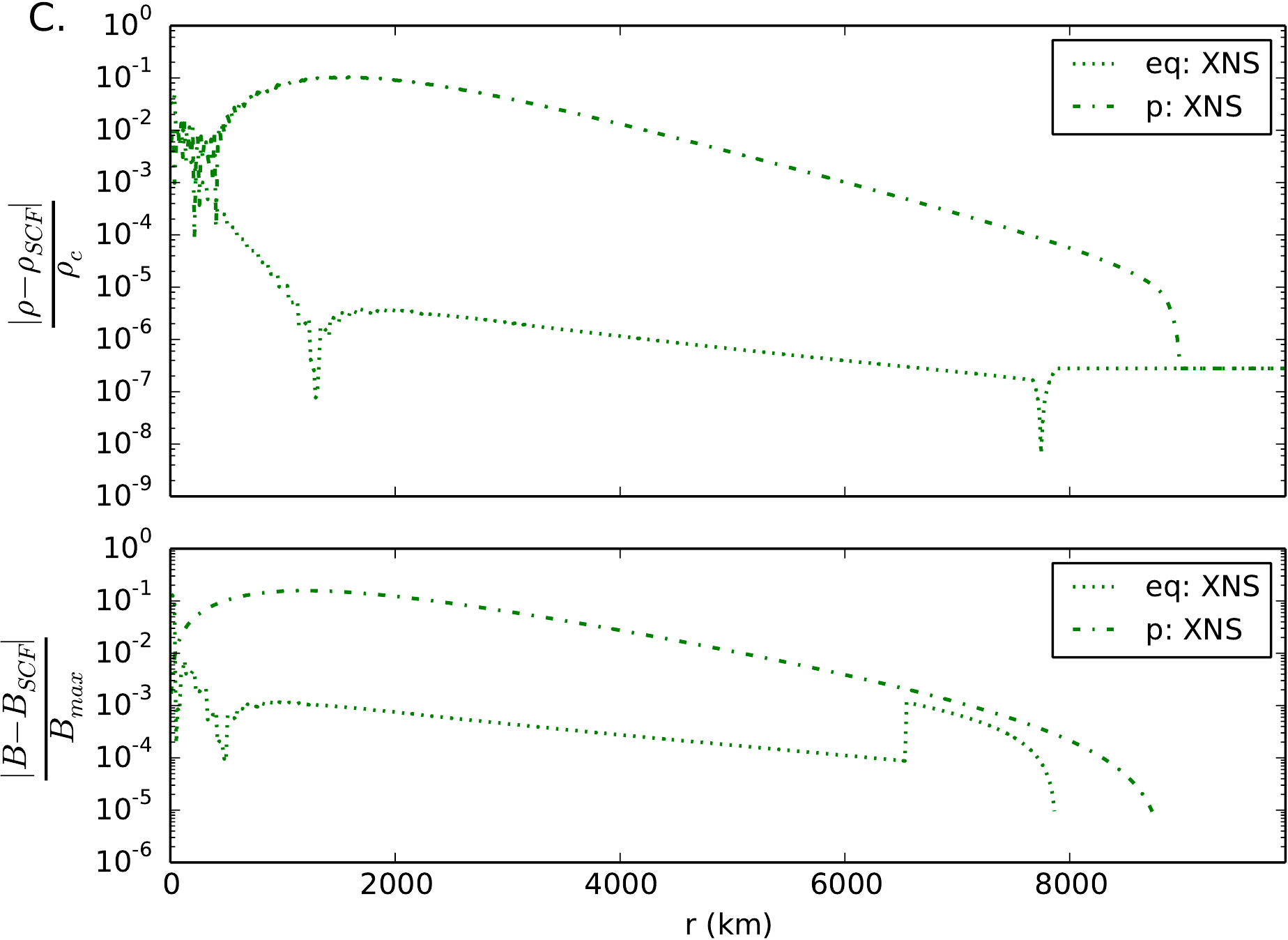}
\caption{\textbf{A.}: Comparison of the density profiles obtained from TOV, Lorene and XNS calculations of the non-magnetic white dwarf configuration presented in Table~\ref{TOV_XNS}. The difference in the density distribution with respect to the TOV solution is displayed here.  The XNS calculation was terminated at a lower density threshold of $10^7$ kg.m$^{-3}$, about $ 10^{-6}$ times the central density. The peak discrepancy between the density profiles is less than 1\%, while the total mass differs by less than 0.1\%, validating the modifications we have made to the Lorene and XNS codes. \textbf{B.}: The difference between the density (upper panel) and field strength (lower panel) distributions obtained from GR and Newtonian calculations of a white dwarf configuration with strong poloidal magnetic field (Table~\ref{poloidal_XNS_Lorene_SCF}). XNS and Lorene results, and polar (p) and equatorial (eq) profiles are shown separately as indicated. Deviations in the profiles due to GR are limited to within $\sim 1$\% of the maximum, while the total mass differs by $\sim 2$\%. \textbf{C.}: Same as panel \textbf{B} but for pure toroidal field (Table \ref{toroidal_XNS_SCF}).  The GR correction here goes up to 10\% of the maximum in the profiles and $\sim 4$\% in total mass.}
\label{XNS_Lorene}
\end{figure}

The XNS and Lorene results using our modifications are compared with the TOV results in Table \ref{TOV_XNS} and Fig \ref{XNS_Lorene}\textbf{A}. As it is seen, the maximum difference in mass is less than 0.1\% and that in radius is less than 1\% for the same central density. The density distribution in the stellar interior matches closely in all three cases.

\subsubsection {Magnetic configurations in General Relativity}

To illustrate the effect of GR on a magnetic white dwarf configuration, we compare in Tables \ref{poloidal_XNS_Lorene_SCF} \& \ref{toroidal_XNS_SCF} the results of GR and Newtonian calculations for a central density 2$\times10^{13}$ kg.m$^{-3}$.   The two tables correspond to poloidal and toroidal field geometries respectively. The poloidal field configuration is constructed by specifying a toroidal current distribution of the form $J^\phi = (e+P) k_{pol}$. We set $k_{pol} = 0.01164$ and $\xi =0$ in XNS to yield a central field strength of 59.6 GT. This combination of central density and field strength yields a near-limiting mass configuration.  In Lorene code the same field strength is achieved by setting the parameter CFA = 950. With these settings the GR calculations yield masses 1.855 M$_\odot$ (XNS) and 1.854 M$_\odot$ (Lorene), with radii of 860 km and 866 km respectively (Table \ref{poloidal_XNS_Lorene_SCF}). Comparison with the corresponding Newtonian configuration shows that the mass of this 
configuration is reduced by $\sim$2\% when general relativistic effects are included. The density and field distributions of the GR and Newtonian configurations differ very little, as seen in Fig. \ref{XNS_Lorene}\textbf{B}.

\begin{figure}
\centering
\includegraphics[width=0.47\textwidth]{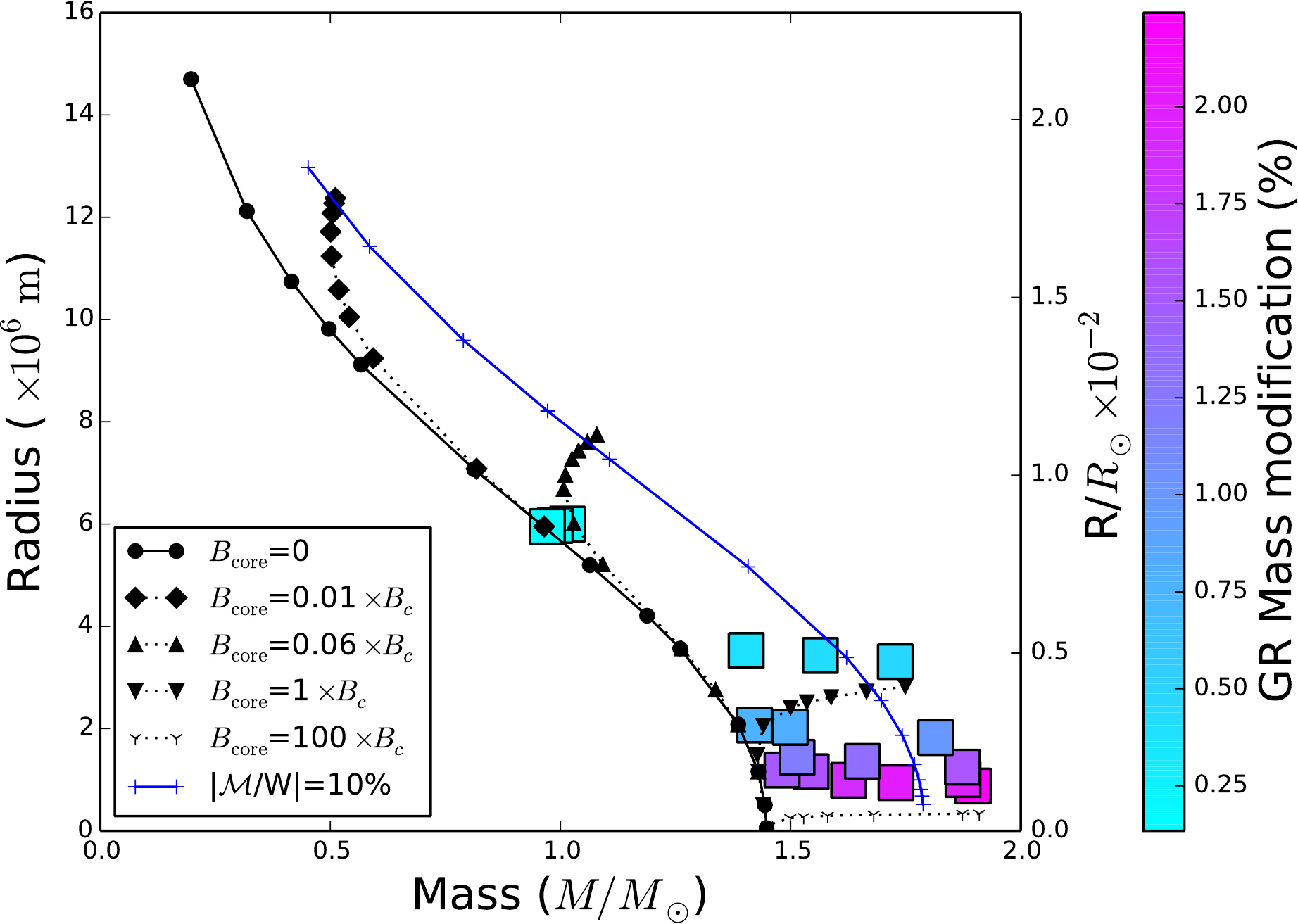}
\caption{Mass-radius relation of white dwarfs with poloidal magnetic field. Lines are derived from Newtonian SCF calculations and symbols along them indicate the assumed central field strength (expressed in units of  $B_c \equiv$ 4.41$\times10^9$~T). Configurations with a fixed central magnetic field start to deviate from the non-magnetic curve as the central density is reduced. Configurations with a fixed ratio of magnetic energy to gravitational energy are described by a mass-radius relation similar to that of non-magnetic ones, but shifted to a higher mass. The curve for $|\mathcal{M}/W| =10$\% is shown, it has a maximum mass of about 1.79 M$_\odot$. The squares indicate the percentage of modification in mass if  general relativistic effects are included.}
\label{poloidal}
\end{figure}

We construct a number of poloidal field configurations with different central densities and field strengths to evaluate the difference in mass derived from Newtonian and GR treatments. The results are shown in Fig. $\ref{poloidal}$. As expected, the difference is larger for more compact configurations. The difference is large also for stronger magnetic fields as they yield more massive configurations.

For a pure toroidal field with $|\mathcal{M}/W|~ \sim$ 39\% and the central density the same as above, the resulting XNS configuration has a mass of 4.95 M$_\odot$, which is $\sim 4$\% smaller than the Newtonian result (Table \ref{toroidal_XNS_SCF}). The radial distribution of density and field agree well in the two cases (Fig. \ref{XNS_Lorene}\textbf{C}).

We have not carried out a similar comparison of GR and Newtonian results for mixed field confgurations, but expect the general trend of reduction in mass by a few percent to apply in this case as well.

\subsection{Salpeter EoS}

So far we have used ideal Fermi degenerate EoS to obtain configurations with different field geometry. We now investigate the consequence of including many body effects in the EoS, \'a la \cite{salpeter61}. Table \ref{poloidal_Salpeter_Lorene} presents a comparison of the pure poloidal field configurations using ideal Fermi EoS, and those using Salpeter EoS for Helium and Carbon compositions respectively. The computations here are carried out in full GR, using the Lorene code. All three configurations have the same central density, 2$\times10^{13}$ kg.m$^{-3}$ and the central magnetic field strength is adjusted to be as close to each other as possible ($\sim$ 59.6 GT). We see that the mass of the configuration is reduced upon inclusion of the many-body effects. The reduction is $\sim$ 0.7\% in case of He composition and $\sim$ 2\% in case of Carbon. In the absence of the magnetic field, the same central density would give masses of 1.403 M$_\odot$ and 1.386 M$_\odot$ with Salpeter EoS, for He and C 
composition respectively. The central density chosen is below the neutronisation threshold in either case. However if the composition is Fe, then the neutronisation threshold is $1.1\times10^{12}$ kg.m$^{-3}$ which yields a maximum stellar mass of 1.45 M$_\odot$ with strong poloidal field.

\begin{table*}
 \begin{tabular}{|c|c|c|c|}
  \hline
    \bf        Code : 	Lorene	&	\bf{Ideal degenerate EoS}	&	\bf{Salpeter EoS for He} &	\bf{Salpeter EoS for C}\\
  \hline \hline
  	\bf CONDITIONS	&			&				&\\

	$\rho_c$	&	2$\times10^{13}$ kg.m$^{-3}$&	2$\times10^{13}$ kg.m$^{-3}$	&	2$\times10^{13}$ kg.m$^{-3}$	\\
	Field parameter	&	CFA = 950 	&	CFA = 954	&	CFA = 959.5\\
	\hline
	\bf   RESULTS	&			&			&\\
	Mass (M$_\odot$)&  	1.8539		&	1.8405		&	1.8194	\\
	R$_{eq}$ (KM)	&  	865.95		&	861.47		&	856.28\\
	R$_p$/R$_{eq}$	&	0.6767		&	0.6752		&	0.6749\\
	B$_{max}$ (GT)	&	59.621		&	59.607		&	59.592\\
	$|\mathcal{M} /W|$&	0.1276		&	0.1284		&	0.1285\\
	Virial test	&	\begin{tabular}[c]{@{}c@{}} $\mid$GRV2$\mid$ = 2.98$\times10^{-8}$\\ $\mid$GRV3$\mid$ = 1.56$\times10^{-8}$\end{tabular}	&	\begin{tabular}[c]{@{}c@{}} $\mid$GRV2$\mid$ = 2.694$\times10^{-7}$\\ $\mid$GRV3$\mid$ = 1.696$\times10^{-7}$\end{tabular}
	&	\begin{tabular}[c]{@{}c@{}} $\mid$GRV2$\mid$ = 2.336$\times10^{-7}$\\ $\mid$GRV3$\mid$ = 2.135$\times10^{-7}$\end{tabular}\\
  \hline
 \end{tabular}
\caption{Pure poloidal field Configurations using ideal degenerate EoS and Salpeter EoS for Helium and Carbon from relativistic code Lorene.}
\label{poloidal_Salpeter_Lorene}
\end{table*}

\section{Conclusions}\label{conclusions}

In this paper, we have computed strongly magnetized axisymmetric white dwarf configurations, considering poloidal, toroidal and mixed poloidal-toroidal field geometries. In all cases, the increase of magnetic energy in a configuration increases its mass. We carry out our calculations in both Newtonian and GR, and quantitatively evaluate the magnitude of corrections introduced by GR effects. For a typical white dwarf, the radius is about a thousand times larger than its Schwarzschild radius, indicating that the GR corrections are expected to be small. From our calculations we find that the correction to the mass of the configuration is at most 2-4\% depending on the field geometry.   Our main results are :

\begin{enumerate}
 \item The Lorentz force has a strong effect on the structure of the stellar configuration. The magnitude and the orientation of the Lorentz force depend on the field geometry. For pure toroidal field configuration the magnetic effects could be extremely large and configurations with mass more than 5 M$_\odot$ can be obtained.
 \item The inclusion of general relativistic effects in the calculation of stellar structure, in general, reduces the stellar mass and radius. The percentage of modification is dependent on the compactness of the configuration. For a pure poloidal field geometry, GR reduces the maximum mass by about 2\% and for a pure toroidal field geometry ithe maximum correction is about 4\%.
 \item The inclusion of many body interaction effects in the EoS also modifies the stellar configuration and reduces the maximum mass. The extent of reduction in mass is dependent on the composition of the white dwarf. For Carbon or heavier atoms, this correction is larger than that due to GR effects.
\end{enumerate}

The configurations presented here satisfy conditions of equilibrium, but are not tested for stability. We restrict the interior density below the neutronization threshold and the GR instability threshold. However the strong magnetic field in the interior may itself drive interchamge instabilities in short time scales \citep{bera_b14}. Further, pure poloidal and pure toroidal magnetic field distributions are not stable by themselves and may evolve over Alf\' ven time scale \citep{marke73, tayler1973, mitchell+2015}. An appropriate mixture of poloidal and toroidal fields may provide stability \citep{braithwaite09, braithwaite06}. The mixed field configurations explored by us have, however, a relatively small toroidal component ($|\mathcal{M}_{tor} /\mathcal{M}| \le$ 7\%), which may not be sufficient to achieve stability. It appears unlikely that mixed field configurations with ($ \ge$ 10\%) magnetic energy in the toroidal component could be constructed in axisymmetry \citep{ciolfi+rezzolla13,armaza+15}, 
and more complex, non-axisymmetric structures may be responsible for the $\ge$ 2 M$_\odot$ white dwarf progenitors of over-luminous Type Ia supernovae.

\section{Acknowledgement}
PB thanks CSIR, India for Research Fellowship grant.

\def\aj{AJ}%
\def\actaa{Acta Astron.}%
\def\araa{ARA\&A}%
\def\apj{ApJ}%
\def\apjl{ApJ}%
\def\apjs{ApJS}%
\def\ao{Appl.~Opt.}%
\def\apss{Ap\&SS}%
\def\aap{A\&A}%
\def\aapr{A\&A~Rev.}%
\def\aaps{A\&AS}%
\def\azh{AZh}%
\def\baas{BAAS}%
\def\bac{Bull. astr. Inst. Czechosl.}%
\def\caa{Chinese Astron. Astrophys.}%
\def\cjaa{Chinese J. Astron. Astrophys.}%
\def\icarus{Icarus}%
\def\jcap{J. Cosmology Astropart. Phys.}%
\def\jrasc{JRASC}%
\def\mnras{MNRAS}%
\def\memras{MmRAS}%
\def\na{New A}%
\def\nar{New A Rev.}%
\def\pasa{PASA}%
\def\pra{Phys.~Rev.~A}%
\def\prb{Phys.~Rev.~B}%
\def\prc{Phys.~Rev.~C}%
\def\prd{Phys.~Rev.~D}%
\def\pre{Phys.~Rev.~E}%
\def\prl{Phys.~Rev.~Lett.}%
\def\pasp{PASP}%
\def\pasj{PASJ}%
\def\qjras{QJRAS}
\def\rmxaa{Rev. Mexicana Astron. Astrofis.}%
\def\skytel{S\&T}%
\def\solphys{Sol.~Phys.}%
\def\sovast{Soviet~Ast.}%
\def\ssr{Space~Sci.~Rev.}%
\def\zap{ZAp}%
\def\nat{Nature}%
\def\iaucirc{IAU~Circ.}%
\def\aplett{Astrophys.~Lett.}%
\def\apspr{Astrophys.~Space~Phys.~Res.}%
\def\bain{Bull.~Astron.~Inst.~Netherlands}%
\def\fcp{Fund.~Cosmic~Phys.}%
\def\gca{Geochim.~Cosmochim.~Acta}%
\def\grl{Geophys.~Res.~Lett.}%
\def\jcp{J.~Chem.~Phys.}%
\def\jgr{J.~Geophys.~Res.}%
\def\jqsrt{J.~Quant.~Spec.~Radiat.~Transf.}%
\def\memsai{Mem.~Soc.~Astron.~Italiana}%
\def\nphysa{Nucl.~Phys.~A}%
\def\physrep{Phys.~Rep.}%
\def\physscr{Phys.~Scr}%
\def\planss{Planet.~Space~Sci.}%
\def\procspie{Proc.~SPIED}%
\let\astap=\aap
\let\apjlett=\apjl
\let\apjsupp=\apjs
\let\applopt=\ao

\bibliographystyle{mnras}	

\bibliography{ref.bib}
\label{lastpage}
 \end{document}